\pdfoutput=1
\documentclass[fleqn,usenatbib]{mnras}
\usepackage[british]{babel}             
\usepackage{amssymb}
\usepackage{amsmath}
\usepackage{txfonts}                      
\usepackage[T1]{fontenc}               
\usepackage{graphicx}                    
\usepackage{url}
\hypersetup{pdfauthor={S.J.Murphy},pdftitle={Gaia's view of the lambda Boo star puzzle}}

\title[Gaia's view of $\lambda$\,Boo stars]{Gaia's view of the $\lambda$\,Boo star puzzle}

\author[Simon J. Murphy \& Ernst Paunzen] 
{Simon J. Murphy$^{1,2}$ \& Ernst Paunzen$^{3}$\\
$^1$Sydney Institute for Astronomy (SIfA), School of Physics, The University of Sydney, NSW 2006, Australia\\
$^2$Stellar Astrophysics Centre, Department of Physics and Astronomy, Aarhus University, DK-8000 Aarhus C, Denmark\\
$^3$Department of Theoretical Physics and Astrophysics, Masaryk University, Kotl\'{a}\v{r}sk\'{a} 2, 611 37 Brno, Czech Republic\\
\\
email: simon.murphy@sydney.edu.au; epaunzen@physics.muni.cz
}
 
\begin{document}

\maketitle 

\begin{abstract}
The evolutionary status of the chemically peculiar class of $\lambda$\,Boo stars has been intensely debated. It is now agreed that the $\lambda$\,Boo phenomenon affects A stars of all ages, from star formation to the terminal age main sequence, but the cause of the chemical peculiarity is still a puzzle. We revisit the debate of their ages and temperatures in order to shed light on the phenomenon, using the new parallaxes in Gaia Data Release 1 with existing Hipparcos parallaxes and multicolour photometry. We find that no single formation mechanism is able to explain all the observations, and suggest that there are multiple channels producing $\lambda$\,Boo spectra. The relative importance of these channels varies with age, temperature and environment.
\end{abstract}


\section{Introduction}
\label{sec:intro}

Chemically peculiar stars enable the study of astrophysical processes from a different perspective. 
At spectral type A there are many classes of chemically peculiar stars, of which the Am and magnetic 
Ap stars are most common. Their slow rotation facilitates investigation of diffusion and mixing processes modulo the presence of a magnetic field. 
The $\lambda$\,Boo stars, which make up about 2\% of the population \citep{gray&corbally1998,paunzen2001}, 
are the third most populous class. Unlike the Am and Ap stars, the $\lambda$\,Boo stars are moderate rotators 
\citep{abt&morrell1995} and the origin of their peculiarity remains a puzzle.

The abundance profile of $\lambda$\,Boo stars is characterised by depletions of Fe-peak elements of up to 2\,dex \citep{andrievskyetal2002}, 
but near-solar abundances of the volatile elements C, N, O and S \citep{baschek&slettebak1988, kampetal2001}. This can 
be explained by the accretion of metal poor material \citep{venn&lambert1990,turcotte&charbonneau1993}. In the circumstellar 
environment, the refractory elements that are incorporated into dust grains are prevented from accreting onto the A star by 
its strong radiation field, while the gas remains to be accreted freely \citep{watersetal1992}. However, the source of the 
accreting material is still uncertain. Suggestions have included accretion of material left-over from star formation 
\citep{holweger&stuerenburg1993}, stars passing through over-dense regions of the interstellar medium (ISM; \citealt{kamp&paunzen2002}), and 
the accretion of material ablated from hot Jupiters \citep{jura2015}. The gravitational influence of a jovian 
companion may further inhibit the migration of dust from a protoplanetary disk to the surface of the star \citep{kamaetal2015}.

Plenty of evidence suggests that $\lambda$\,Boo stars are young. They have been found in OB associations \citep{gray&corbally1993}, 
they are absent from intermediate-age open clusters despite extensive searches \citep{gray&corbally2002}, $\lambda$\,Boo abundance 
patterns are seen in several pre-main-sequence stars \citep{folsometal2012}, and luminosities from \textit{Hipparcos} parallaxes 
place some $\lambda$\,Boo stars very close to the zero-age main sequence \citep{paunzen1997}. However, arguments to the contrary can 
also be found. \citet{iliev&barzova1995} found they occupy a range of ages between $10^7$ and $10^9$\,yr, and \citet{ilievetal2002} 
found a binary pair of $\lambda$\,Boo stars with an age of $10^9$\,yr, ruling out a pre-main sequence status. A continuous source of 
accreting material must therefore be present, supplying $10^{-12}$ to $10^{-11}$\,M$_{\odot}$yr$^{-1}$ of material \citep{turcotte2002}, 
else the peculiarities are erased within $10^6$\,yr \citep{turcotte&charbonneau1993}.

Attempts have been made to explain the $\lambda$\,Boo class as a heterogeneous group of stars that are not intrinsically chemically peculiar, but have composite spectra due to binarity, instead \citep{faraggiana&bonifacio1999}. Within the `composite spectrum hypothesis', the metal lines of one star are veiled by the continuum of the other and vice-versa, such that all metal lines {\it appear} peculiarly weak, but aren't. The hypothesis was advocated in a series of papers by the same authors \citep[e.g.][]{faraggianaetal2001b}, but was refuted in critical independent analyses \citep{stutz&paunzen2006,griffinetal2012}. 

A significant fraction of $\lambda$ Boo stars are located within the classical $\delta$ Sct/$\gamma$ Dor 
instability strip. \citet{paunzenetal2002a} presented a detailed analysis of their pulsational behaviour, based on ground-based photometry. 
They concluded that at least 70\% of the group members inside the classical instability strip pulsate, 
and they do so with high overtone p-modes (Q\,$<$\,0.020\,d). This group of stars is perfectly suited to test pulsational models
of stars with non-solar atmospheric composition.
Recent advances in determining precise main-sequence ages of $\gamma$\,Dor stars from asteroseismology offer a reciprocal benefit \citep{kurtzetal2014,saioetal2015}.

An opportunity to re-evaluate the evolutionary status of $\lambda$\,Boo stars has arrived with {\it Gaia} data 
release 1 (DR1, \citealt{gaiacollaboration2016,lindgreenetal2016}), which has added heavily to the number of $\lambda$\,Boo stars with precise parallaxes. We examine an updated HR diagram of these stars in an attempt to decipher the enigma surrounding their peculiarities. In Sect.\,\ref{sec:extraction} we describe our data extraction procedures. The HR diagram is described in Sect.\,\ref{sec:HRD} and interpreted in Sect.\,\ref{sec:discussion}, then brief conclusions are given in Sect.\,\ref{sec:conclusions}.

\section{Data extraction}
\label{sec:extraction}

The starting point of our target selection was the catalogue of bona-fide $\lambda$\,Boo stars by \citet{murphyetal2015b}. From this sample, we made two selections.
Firstly, we selected objects for which parallaxes with an error smaller than 25\% in the Hipparcos catalogue \citep{vanleeuwen2007} and/or 
Gaia DR1 \citep{gaiacollaboration2016} are available. Most of the parallaxes are much smaller than 25\%, as we discuss in Sect.\,\ref{ssec:pi}.
Secondly, we included only those stars with available Johnson $UBV$, Str{\"o}mgren-Crawford $uvby\beta$, and Geneva 7-colour photometry, taken from \citet{paunzen2015} and the General Catalogue of Photometric Data
(GCPD\footnote{http://gcpd.physics.muni.cz/}). Where possible, averaged and weighted mean values were used throughout. The final sample consists of 172 stars.

\subsection{Colours and reddening}
\label{ssec:colours}

Throughout this analysis, the following relation between extinction, $A$, and colour excess (reddening), $E$, in specific bands, is used: 
$$A_V=3.1E_{(B-V)}=4.3E_{(b-y)}=4.95E_{(B2-V1)}.$$
The reddening for the targets was estimated using photometric calibrations in the
Str{\"o}mgren $uvby\beta$ \citep{crawford1978,crawford1979} and the $Q$-parameter within
the Johnson $UBV$ system \citep{johnson1958}. These methods are based only on photometric indices and do not take
any distance estimates via parallax measurements into account. In addition, for all stars, the distance (derived from the parallaxes) and Galactic 
coordinates were used to determine the reddening from the \citet{amores&lepine2005} extinction model. If several estimates 
were available, they were compared and were always in excellent agreement. 

The effective temperatures ($T_{\rm eff}$) of the individual stars were calibrated by using the measurements in the
Johnson $UBV$, Geneva 7-colour, and Str{\"o}mgren-Crawford $uvby\beta$ photometric systems \citep{golay1974}.
This approach was chosen because it yields the most homogeneous result; taking $T_{\rm eff}$ values
from the literature, e.g. from spectroscopy, introduces a variety of unknown biases \citep{lebzelteretal2012}, such as
the use of different stellar atmosphere models, spectral resolutions, analytic methods and so on. Here, we used the following calibrations for the three photometric systems:

\begin{itemize}
\item Johnson $UBV$: \citet{paunzenetal2005b,paunzenetal2006}
\item Geneva 7-colour: \citet{kunzlietal1997,paunzenetal2005b,paunzenetal2006} 
\item Str{\"o}mgren--Crawford $uvby\beta$: \citet{moon&dworetsky1985,napiwotzkietal1993,balona1994,ribasetal1997,paunzenetal2005b,paunzenetal2006}
\end{itemize}

The individual $T_{\rm eff}$ values within each photometric system were first checked
for intrinsic consistency and then averaged. The error of the photometric measurements in
the used systems for our targets are normally below 0.01\,mag. \citet{kunzlietal1997} investigated
in detail the influence of the observational errors on the $T_{\rm eff}$ estimates, finding the dispersion and systematic zero-point shifts to be small. 
These final $T_{\rm eff}$ values,
together with the standard deviations of the means, are listed in Table \ref{table_parameters}. Also given there are the absolute and apparent magnitudes ($M_{\rm V}$, $m_{\rm V}$) of each star, calculated using their parallax measurements ($\pi$) by applying the basic formula
\begin{equation}
M_\mathrm{V} = m_\mathrm{V} + 5\,({\rm log}\,\pi + 1) - A_\mathrm{V}.
\end{equation}
We used the $m_\mathrm{V}$ values given by \citet{kharchenko2001}, who transformed the Tycho data uniformly to the Johnson system. This is the most
homogeneous sample available for this data type.

Finally, the bolometric correction (B.C.) for a given $T_{\rm eff}$ was derived using the relation by \citet{flower1996}.

For the error estimate of $\log L$, only the error of the parallax was taken into account.
The mean error of the mean extinctions ($\overline{\sigma_A}$) is 0.01\,mag, which propagates to an error contribution
of 0.004\,dex for the final luminosities. The B.C. for A stars is very flat. It ranges from $+$0.03\,mag to $-$0.06\,mag for 7000\,$<$\,$T_{\rm eff}$\,$<$\,9000\,K. Even a $T_{\rm eff}$ error of 30\%  
results in a B.C. error contribution to the luminosity of 0.006\,dex.

\subsection{Parallax validation}
\label{ssec:pi}

Where both Gaia and Hipparcos parallaxes were available, we used the measurement with the smallest uncertainty.

\begin{figure*}
\begin{center}
\includegraphics[width=0.9269\textwidth]{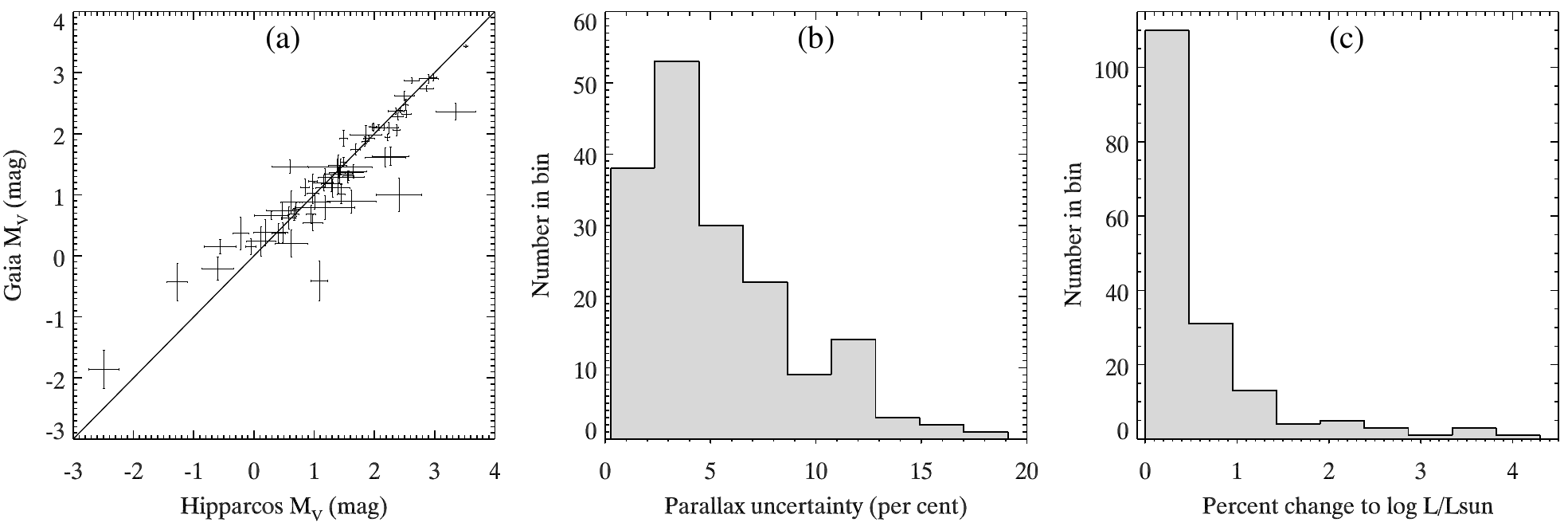}
\caption{{\bf (a)} A comparison of the absolute magnitudes derived from the parallaxes of Hipparcos and Gaia DR1, showing no obvious offset. The diagonal line represents $M_{\rm V,Gaia} = M_{\rm V,Hip}$. {\bf (b)} The distribution of fractional parallax uncertainties, $f$. Where both Gaia and Hipparcos parallaxes were available, we used the one having the smaller uncertainty; 86\% of our sample have $f$\,$<$\,$10$\%. {\bf (c)} Distribution of the percentage change in $\log L/L_{\sun}$ that would result if a Lutz-Kelker correction (Eq.\,\ref{eq:LK}) were applied. We applied no such correction. Note that the vertical scales of {\bf (b)} and {\bf (c)} differ.}
\label{fig:mvs}
\end{center}
\end{figure*}

Figure \ref{fig:mvs}a shows the comparison of the absolute magnitudes derived from the Hipparcos and Gaia DR1, respectively.
No significant offset for the 69 stars in common has been detected: $\Delta \overline{M_\mathrm{V}}$\,=\,+0.05$\pm$0.41\,mag.

Three stars (HD\,21335, HD\,105058, and HD\,114930) were found to have a $\Delta \overline{M_\mathrm{V}}$ value that is 
$2\sigma$ larger than calculated from the individual errors. From these, HD\,21335 and HD\,105058
are listed as visual and/or spectroscopic binary
systems in the Washington Visual Double Star Catalog \citep[WDS,][]{masonetal2001} and/or by \citet{faraggianaetal2004}. 
Parallax measurements for binaries already encountered difficulties in the Hipparcos era \citep{masonetal1999}, 
with very erratic parallaxes being obtained for several systems, 
depending on the separation and magnitude difference of the components. The behaviour of the Gaia
measurements in this respect is still to be investigated in detail. For 
HD\,114930 ($\Delta M_\mathrm{V}$\,=\,+0.84\,mag), we have found no plausible explanation for the large differences. Since for this object the Gaia parallax has the smaller fractional parallax uncertainty [$f_{\rm G} = \sigma_{\rm G}(\pi)/\pi = 6.7$\%], this is the one we used. The Hipparcos fractional parallax uncertainty is unusually large ($f_{\rm H} = 15.3$\%). The distribution of $f$ is shown in Fig.\,\ref{fig:mvs}b.

A systematic bias occurs when using parallax-limited samples \citep{trumpler&weaver1953}, 
which on average leads to the over-estimation of stellar distances. 
\citet{lutz&kelker1973} calculated corrections for the resulting bias
in the absolute magnitudes of individual stars. 
Their correction reaches 0.43\,mag for $f$\,=\,17.5\%.
However, since the publication
of the Hipparcos data, a debate has emerged on the appropriateness of this correction 
for single stars \citep{smith2003,francis2014}. 
On the basis of Monte-Carlo simulations, \citet{francis2014} showed that the overall (net) correction for the absolute magnitude can be described as 
\begin{equation}
\label{eq:LK}
\Delta(M_{\mathrm V}) = -5.35\left(\frac{\sigma(\pi)}{\pi}\right)^{2},
\end{equation}
but he also wrote ``there is, in practice, no circumstance in
which the [Lutz-Kelker] correction should be applied.''

That is not to say that no bias exists, rather, that the Lutz-Kelker correction is not the appropriate treatment. 
A full Bayesian treatment is preferred but not straightforward, especially for \mbox{$f$\,$>$\,$0.2$}, 
since it requires the difficult selection of an appropriate prior \citep{bailer-jones2015}.

In this work, we have calculated the numerical correction for each star according to Eq.\,(\ref{eq:LK}) to evaluate its significance. 
We found the resulting correction to our logarithmic luminosities to be smaller than 1\% 
for 82\% of our sample (Fig.\,\ref{fig:mvs}c). In no case does this correction exceed 
the 1$\sigma$ uncertainty in $\log L$. We therefore deemed the correction insignificant, and made no correction to our data.

It is useful to compare literature estimates of the significance of bias corrections for similar stellar populations. 
The sample of 90 Mount Wilson Subgiants \citep{adamsetal1935} is particularly useful here, 
having comparable absolute magnitudes, dispersion, and fractional parallax uncertainties to our $\lambda$\,Boo stars. 
\citet{sandageetal2016} applied the \citet{sandage&saha2002} treatment of bias to the Mount Wilson Subgiants, 
and also found the resulting corrections to be insignificant.

\begin{table*}
\begin{center}
\caption{Parameters for the 172 $\lambda$\,Boo stars and candidates considered in this work, with Hipparcos (`Hip') or Gaia parallaxes. Membership probability (`mem.') in the $\lambda$\,Boo class (from \citealt{murphyetal2015b}) is indicated with integers between 1 and 4, with 1 being certain members and 4 being non-members. B.C. stands for bolometric correction, and other symbols have their usual usage as described in the text. Where absolute magnitudes are missing, no parallax was available in that survey. The final column gives the variability type, from sources described in Sect.\,\ref{ssec:pulsation}.}
\label{table_parameters}
\begin{tabular}{cccrcrr@{ $\pm$ }rr@{ $\pm$ }rr@{ $\pm$ }rr@{ $\pm$ }rc}
\hline
\hline
HD & HIP/TYC & \hspace{-2mm}mem.\hspace{-2mm} & $m_\mathrm{V}$ & $A_\mathrm{V}$ & B.C. & \multicolumn{2}{c}{$T_\mathrm{eff}$} & \multicolumn{2}{c}{$M_\mathrm{V,Hip}$} & \multicolumn{2}{c}{$M_\mathrm{V,Gaia}$}  & \multicolumn{2}{c}{$\log L/L_{\sun}$}  & var? \\
& & & (mag) & (mag) & (mag) & \multicolumn{2}{c}{(K)} & \multicolumn{2}{c}{(mag)} & \multicolumn{2}{c}{(mag)} & \multicolumn{2}{c}{} & \\
\hline
3	&	424	&	3	&	6.685	&	0.000	&	+0.01	&	8390	&	232	&	+0.98	&	0.16	&	+0.54	&	0.13	&	1.69	&	0.05	&	---	\\
319	&	636	&	1	&	5.931	&	0.012	&	+0.02	&	8080	&	77	&	+1.45	&	0.06	&	+1.01	&	0.16	&	1.32	&	0.03	&	const.	\\
2904	&	2628	&	3	&	6.406	&	0.023	&	$-$0.21	&	9820	&	247	&	+0.48	&	0.09	&	+0.38	&	0.16	&	1.71	&	0.04	&	---	\\
4158	&	5849--1699--1 &		1	&	9.536	&	0.577	&	+0.03	&	7820	&	50	&		\multicolumn{2}{c}{}		&	+0.88	&	0.27	&	1.55	&	0.11	&	---	\\
6173	&	3271--1475--1	&	3	&	8.520	&	0.490	&	$-$0.25	&	9990	&	31	&		\multicolumn{2}{c}{}		&	$-$0.46	&	0.31	&	2.09	&	0.13	&	---	\\
6870	&	5321	&	1	&	7.473	&	0.000	&	+0.03	&	7300	&	87	&	+2.39	&	0.09	&	+2.28	&	0.06	&	0.99	&	0.02	&	$\delta$\,Sct	\\
7908	&	6108	&	1	&	7.304	&	0.000	&	+0.03	&	7120	&	73	&	+2.87	&	0.11	&	+2.74	&	0.05	&	0.81	&	0.02	&	const.	\\
9100	&	6981	&	4	&	6.013	&	0.000	&	+0.03	&	7760	&	88	&	+0.19	&	0.20	&	+0.38	&	0.22	&	1.82	&	0.08	&	$\delta$\,Sct	\\
11413	&	8593	&	1	&	5.933	&	0.007	&	+0.02	&	7870	&	82	&	+1.49	&	0.05	&	+1.53	&	0.08	&	1.30	&	0.02	&	$\delta$\,Sct	\\
11503	&	8832	&	4	&	4.598	&	0.018	&	$-$0.35	&	10480	&	193	&	+1.07	&	0.10	&		\multicolumn{2}{c}{}		&	1.47	&	0.04	&	other	\\
13755	&	10304	&	1	&	7.829	&	0.000	&	+0.03	&	6990	&	86	&	+1.65	&	0.22	&	+1.38	&	0.11	&	1.35	&	0.05	&	$\delta$\,Sct	\\
15164	&	635--338--1	&	1	&	8.281	&	0.000	&	+0.03	&	7120	&	73	&		\multicolumn{2}{c}{}		&	+2.77	&	0.10	&	0.79	&	0.04	&	---	\\
16811	&	12640	&	4	&	5.735	&	0.060	&	$-$0.28	&	10150	&	232	&	+0.34	&	0.09	&		\multicolumn{2}{c}{}		&	1.77	&	0.04	&	---	\\
16955	&	12744	&	4	&	6.376	&	0.000	&	0.00	&	8450	&	164	&	+1.29	&	0.17	&	+1.18	&	0.13	&	1.43	&	0.05	&	---	\\
17138	&	13133	&	2	&	6.262	&	0.042	&	+0.01	&	8320	&	65	&	+2.16	&	0.05	&	+2.10	&	0.05	&	1.06	&	0.02	&	other	\\
21335	&	16077	&	4	&	6.530	&	0.001	&	+0.03	&	7810	&	74	&	+1.09	&	0.14	&	$-$0.41	&	0.33	&	2.07	&	0.13	&	---	\\
22470	&	16803	&	4	&	5.236	&	0.000	&	$-$1.13	&	14350	&	238	&	$-$0.63	&	0.17	&		\multicolumn{2}{c}{}		&	2.15	&	0.07	&	other	\\
23258	&	17453	&	1	&	6.076	&	0.000	&	$-$0.13	&	9380	&	168	&	+1.62	&	0.07	&		\multicolumn{2}{c}{}		&	1.25	&	0.03	&	---	\\
23392	&	17462	&	1	&	8.246	&	0.163	&	$-$0.28	&	10140	&	169	&		\multicolumn{2}{c}{}		&	+1.65	&	0.17	&	1.24	&	0.07	&	const.	\\
24472	&	18153	&	2	&	7.097	&	0.000	&	+0.02	&	6810	&	90	&	+2.36	&	0.13	&	+2.37	&	0.05	&	0.95	&	0.02	&	const.	\\
24712	&	18339	&	4	&	5.984	&	0.000	&	+0.03	&	7240	&	74	&	+2.52	&	0.04	&	+2.47	&	0.09	&	0.89	&	0.02	&	other	\\
27404	&	20262	&	4	&	7.931	&	0.127	&	+0.03	&	7590	&	114	&	+1.26	&	0.47	&	+1.51	&	0.28	&	1.29	&	0.11	&	other	\\
30422	&	22192	&	1	&	6.173	&	0.000	&	0.02	&	7930	&	72	&	+2.43	&	0.04	&	+2.38	&	0.04	&	0.95	&	0.02	&	$\delta$\,Sct	\\
30739	&	22509	&	4	&	4.348	&	0.073	&	$-$0.14	&	9430	&	264	&	+0.09	&	0.06	&		\multicolumn{2}{c}{}		&	1.87	&	0.02	&	---	\\
31293	&	22910	&	4	&	7.064	&	0.035	&	+0.02	&	8170	&	249	&	+1.31	&	0.29	&	+1.11	&	0.15	&	1.46	&	0.06	&	other	\\
31295	&	22845	&	1	&	4.646	&	0.088	&	$-$0.05	&	8900	&	231	&	+1.80	&	0.02	&		\multicolumn{2}{c}{}		&	1.18	&	0.01	&	const.	\\
34787	&	25197	&	4	&	5.236	&	0.058	&	$-$0.23	&	9910	&	65	&	+0.11	&	0.07	&		\multicolumn{2}{c}{}		&	1.86	&	0.03	&	---	\\
34797	&	24827	&	4	&	6.505	&	0.039	&	$-$1.05	&	13860	&	176	&	$-$0.48	&	0.23	&		\multicolumn{2}{c}{}		&	2.09	&	0.09	&	other	\\
35242	&	25205	&	1	&	6.330	&	0.000	&	+0.02	&	8230	&	59	&	+1.69	&	0.08	&	+1.74	&	0.09	&	1.22	&	0.03	&	$\delta$\,Sct	\\
36496	&	26410	&	3	&	6.302	&	0.016	&	0.03	&	7640	&	107	&	+1.83	&	0.09	&		\multicolumn{2}{c}{}		&	1.17	&	0.04	&	---	\\
37886	&	4771--343--1	&	4	&	9.005	&	0.197	&	$-$0.78	&	12490	&	54	&		\multicolumn{2}{c}{}		&	+0.96	&	0.24	&	1.52	&	0.10	&	---	\\
38545	&	27316	&	4	&	5.717	&	0.000	&	+0.02	&	8250	&	218	&	+0.63	&	0.19	&		\multicolumn{2}{c}{}		&	1.65	&	0.08	&	const.	\\
39283	&	27949	&	4	&	4.960	&	0.108	&	$-$0.08	&	9100	&	211	&	0.53	&	0.04	&		\multicolumn{2}{c}{}		&	1.69	&	0.02	&	---	\\
39421	&	27713	&	3	&	5.954	&	0.000	&	+0.02	&	8240	&	100	&	+0.98	&	0.10	&	+1.02	&	0.17	&	1.49	&	0.07	&	---	\\
40588	&	28517	&	1	&	6.182	&	0.114	&	$-$0.03	&	8750	&	167	&	+1.78	&	0.07	&		\multicolumn{2}{c}{}		&	1.19	&	0.03	&	---	\\
42503	&	29159	&	1	&	7.434	&	0.000	&	+0.03	&	7460	&	109	&	+0.47	&	0.27	&	+0.74	&	0.14	&	1.60	&	0.05	&	$\delta$\,Sct	\\
47152	&	31737	&	4	&	5.743	&	0.129	&	$-$0.37	&	10560	&	281	&	+0.48	&	0.14	&		\multicolumn{2}{c}{}		&	1.71	&	0.06	&	other	\\
54272	&	1353--663--1	&	3	&	8.774	&	0.167	&	+0.03	&	6890	&	69	&		\multicolumn{2}{c}{}		&	+2.51	&	0.09	&	0.90	&	0.04	&	$\gamma$\,Dor	\\
56405	&	35180	&	4	&	5.445	&	0.015	&	$-$0.04	&	8830	&	185	&	+0.85	&	0.07	&	+1.12	&	0.14	&	1.56	&	0.03	&	other	\\
64491	&	38723	&	1	&	6.221	&	0.000	&	0.03	&	7110	&	125	&	+2.37	&	0.06	&	+2.06	&	0.09	&	0.95	&	0.03	&	$\delta$\,Sct	\\
66920	&	39041	&	4	&	6.320	&	0.046	&	+0.02	&	8200	&	89	&	+0.70	&	0.06	&	+0.76	&	0.11	&	1.62	&	0.02	&	const.	\\
68758	&	40211	&	4	&	6.518	&	0.000	&	$-$0.04	&	8820	&	132	&	+0.95	&	0.08	&	+0.68	&	0.15	&	1.52	&	0.03	&	---	\\
73210	&	42327	&	4	&	6.735	&	0.037	&	+0.03	&	7710	&	84	&	+0.12	&	0.24	&	+0.24	&	0.24	&	1.80	&	0.09	&	---	\\
73872	&	1395--2222--1	&	4	&	8.317	&	0.000	&	+0.03	&	7790	&	76	&		\multicolumn{2}{c}{}		&	+2.18	&	0.25	&	1.03	&	0.10	&	$\delta$\,Sct	\\
74873	&	43121	&	1	&	5.883	&	0.449	&	$-$0.30	&	10220	&	471	&	+1.77	&	0.05	&		\multicolumn{2}{c}{}		&	1.19	&	0.02	&	const.	\\
74911	&	43134	&	4	&	8.515	&	0.000	&	+0.02	&	7960	&	151	&	+2.18	&	0.34	&	+1.61	&	0.16	&	1.26	&	0.06	&	---	\\
75654	&	43354	&	1	&	6.372	&	0.000	&	+0.03	&	7270	&	85	&	+1.98	&	0.06	&	+2.11	&	0.05	&	1.06	&	0.02	&	$\delta$\,Sct	\\
78316	&	44798	&	4	&	5.236	&	0.059	&	$-$0.98	&	13500	&	117	&	$-$0.88	&	0.09	&		\multicolumn{2}{c}{}		&	2.25	&	0.04	&	other	\\
78661	&	44984	&	4	&	6.474	&	0.000	&	+0.03	&	6830	&	79	&	+3.52	&	0.03	&	+3.43	&	0.02	&	0.53	&	0.01	&	other	\\
79025	&	44979	&	4	&	6.473	&	0.023	&	+0.03	&	7830	&	107	&	+0.41	&	0.11	&	+0.37	&	0.16	&	1.74	&	0.04	&	const.	\\
79108	&	45167	&	4	&	6.132	&	0.048	&	$-$0.22	&	9840	&	259	&	+1.10	&	0.08	&		\multicolumn{2}{c}{}		&	1.46	&	0.03	&	---	\\
79469	&	45336	&	4	&	3.878	&	0.000	&	$-$0.37	&	10550	&	346	&	+1.17	&	0.12	&		\multicolumn{2}{c}{}		&	1.43	&	0.05	&	---	\\
80081	&	45688	&	4	&	3.890	&	0.000	&	+0.01	&	8430	&	219	&	+0.98	&	0.06	&		\multicolumn{2}{c}{}		&	1.51	&	0.03	&	---	\\
81290	&	46011	&	2	&	8.864	&	0.401	&	+0.03	&	7140	&	645	&		\multicolumn{2}{c}{}		&	+1.69	&	0.11	&	1.22	&	0.04	&	const.	\\
82573	&	46813	&	4	&	5.736	&	0.000	&	+0.02	&	8130	&	107	&	+0.42	&	0.09	&		\multicolumn{2}{c}{}		&	1.73	&	0.04	&	const.	\\
83041	&	47018	&	2	&	8.789	&	0.659	&	+0.03	&	7630	&	930	&	+1.18	&	0.50	&	+0.79	&	0.19	&	1.58	&	0.08	&	$\delta$\,Sct	\\
83277	&	47155	&	2	&	8.281	&	0.479	&	+0.03	&	7380	&	690	&	+1.56	&	0.27	&	+1.29	&	0.10	&	1.38	&	0.04	&	const.	\\
84948	&	48243	&	2	&	8.116	&	0.000	&	+0.02	&	6780	&	64	&	+1.62	&	0.41	&	+0.89	&	0.19	&	1.55	&	0.08	&	$\delta$\,Sct	\\
87271	&	49328	&	1	&	7.139	&	0.785	&	$-$0.40	&	10680	&	254	&	+0.29	&	0.28	&	+0.66	&	0.07	&	1.64	&	0.03	&	$\delta$\,Sct	\\
\hline
\end{tabular}
\end{center}
\end{table*}

\addtocounter{table}{-1}

\begin{table*}
\begin{center}
\caption{continued.}
\begin{tabular}{cccrcrr@{ $\pm$ }rr@{ $\pm$ }rr@{ $\pm$ }rr@{ $\pm$ }rc}
\hline
\hline
HD & HIP/TYC & \hspace{-2mm}mem.\hspace{-2mm} & $m_\mathrm{V}$ & $A_\mathrm{V}$ & B.C. & \multicolumn{2}{c}{$T_\mathrm{eff}$} & \multicolumn{2}{c}{$M_\mathrm{V,Hip}$} & \multicolumn{2}{c}{$M_\mathrm{V,Gaia}$}  & \multicolumn{2}{c}{$\log L/L_{\sun}$}  & var? \\
& & & (mag) & (mag) & (mag) & \multicolumn{2}{c}{(K)} & \multicolumn{2}{c}{(mag)} & \multicolumn{2}{c}{(mag)} & \multicolumn{2}{c}{} & \\
\hline
87696	&	49593	&	4	&	4.479	&	0.017	&	+0.02	&	7950	&	76	&	+2.21	&	0.01	&		\multicolumn{2}{c}{}		&	1.02	&	0.00	&	$\delta$\,Sct	\\
89239	&	50459	&	4	&	6.525	&	0.036	&	$-$0.36	&	10500	&	235	&	+0.71	&	0.13	&		\multicolumn{2}{c}{}		&	1.62	&	0.05	&	---	\\
91130	&	51556	&	1	&	5.902	&	0.000	&	+0.02	&	8130	&	152	&	+1.44	&	0.05	&		\multicolumn{2}{c}{}		&	1.32	&	0.02	&	const.	\\
97411	&	54742	&	4	&	6.098	&	0.076	&	$-$0.30	&	10250	&	208	&	+0.09	&	0.24	&		\multicolumn{2}{c}{}		&	1.87	&	0.09	&	---	\\
97773	&	54981	&	3	&	7.552	&	0.021	&	+0.03	&	7420	&	104	&	+0.91	&	0.34	&		\multicolumn{2}{c}{}		&	1.54	&	0.14	&	---	\\
97937	&	55033	&	3	&	6.649	&	0.000	&	+0.03	&	7190	&	76	&	+2.53	&	0.08	&	+2.32	&	0.06	&	0.97	&	0.02	&	---	\\
98353	&	55266	&	4	&	4.755	&	0.000	&	+0.01	&	8410	&	184	&	+0.91	&	0.04	&		\multicolumn{2}{c}{}		&	1.54	&	0.02	&	other	\\
98772	&	55564	&	4	&	6.021	&	0.000	&	+0.01	&	8330	&	156	&	+1.33	&	0.06	&		\multicolumn{2}{c}{}		&	1.37	&	0.03	&	---	\\
100546	&	56379	&	3	&	6.696	&	0.197	&	$-$0.40	&	10720	&	193	&	+1.57	&	0.09	&	+1.31	&	0.08	&	1.38	&	0.03	&	other	\\
100740	&	56553	&	4	&	6.563	&	0.000	&	+0.02	&	7900	&	132	&	+1.42	&	0.09	&	+1.34	&	0.14	&	1.33	&	0.04	&	---	\\
101108	&	56768	&	2	&	8.847	&	0.023	&	+0.03	&	7800	&	75	&		\multicolumn{2}{c}{}		&	+1.32	&	0.22	&	1.37	&	0.09	&	const.	\\
102541	&	57567	&	2	&	7.946	&	0.088	&	+0.03	&	7690	&	108	&	+2.50	&	0.17	&	+2.62	&	0.08	&	0.85	&	0.03	&	$\delta$\,Sct	\\
105058	&	58992	&	1	&	8.873	&	0.026	&	+0.03	&	7670	&	113	&	+2.41	&	0.38	&	+1.00	&	0.27	&	1.50	&	0.11	&	$\delta$\,Sct	\\
105199	&	2527--707--1	&	4	&	9.815	&	0.000	&	+0.03	&	7750	&	117	&		\multicolumn{2}{c}{}		&	+2.47	&	0.41	&	0.91	&	0.17	&	---	\\
105260	&	59107	&	3	&	9.180	&	0.000	&	+0.03	&	7260	&	89	&		\multicolumn{2}{c}{}		&	+2.28	&	0.13	&	0.99	&	0.05	&	---	\\
106223	&	59594	&	1	&	7.416	&	0.313	&	+0.03	&	7000	&	783	&	+1.65	&	0.17	&	+1.37	&	0.09	&	1.35	&	0.04	&	const.	\\
107223	&	60137	&	4	&	8.187	&	0.359	&	$-$0.15	&	9490	&	575	&		\multicolumn{2}{c}{}		&	+0.65	&	0.18	&	1.64	&	0.07	&	const.	\\
107233	&	60134	&	1	&	7.361	&	0.000	&	+0.03	&	7220	&	76	&	+2.62	&	0.12	&	+2.87	&	0.05	&	0.75	&	0.02	&	---	\\
108283	&	60697	&	4	&	4.910	&	0.155	&	+0.03	&	7140	&	50	&	+0.12	&	0.04	&		\multicolumn{2}{c}{}		&	1.85	&	0.02	&	---	\\
108714	&	60933	&	4	&	7.718	&	0.000	&	+0.01	&	8310	&	329	&	+1.40	&	0.23	&	+1.34	&	0.31	&	1.34	&	0.09	&	---	\\
108765	&	60957	&	4	&	5.671	&	0.097	&	$-$0.02	&	8720	&	227	&	+0.95	&	0.05	&		\multicolumn{2}{c}{}		&	1.52	&	0.02	&	---	\\
109738	&	9228--358--1	&	1	&	8.284	&	0.086	&	+0.03	&	7600	&	102	&		\multicolumn{2}{c}{}		&	+2.17	&	0.10	&	1.03	&	0.04	&	$\delta$\,Sct	\\
109980	&	61692	&	4	&	6.346	&	0.000	&	+0.03	&	7670	&	120	&	+2.19	&	0.06	&		\multicolumn{2}{c}{}		&	1.02	&	0.03	&	other	\\
110377	&	61937	&	4	&	6.212	&	0.000	&	+0.03	&	7700	&	80	&	+2.22	&	0.04	&	+1.94	&	0.05	&	1.01	&	0.02	&	$\delta$\,Sct	\\
110411	&	61960	&	1	&	4.864	&	0.063	&	$-$0.06	&	8970	&	112	&	+2.00	&	0.02	&		\multicolumn{2}{c}{}		&	1.10	&	0.01	&	const.	\\
111005	&	62318	&	2	&	7.962	&	0.077	&	+0.02	&	6790	&	65	&	+1.86	&	0.26	&	+1.98	&	0.15	&	1.11	&	0.06	&	const.	\\
111164	&	62394	&	4	&	6.096	&	0.000	&	+0.02	&	8210	&	77	&	+1.49	&	0.07	&	+1.92	&	0.13	&	1.30	&	0.03	&	---	\\
111604	&	62641	&	1	&	5.875	&	0.000	&	+0.03	&	7640	&	105	&	+0.67	&	0.08	&	+0.68	&	0.10	&	1.63	&	0.03	&	$\delta$\,Sct	\\
111786	&	62788	&	1	&	6.136	&	0.017	&	+0.03	&	7490	&	100	&	+1.99	&	0.05	&	+2.11	&	0.07	&	1.10	&	0.02	&	$\delta$\,Sct	\\
111893	&	62825	&	4	&	6.290	&	0.000	&	+0.03	&	7710	&	92	&	+1.07	&	0.10	&		\multicolumn{2}{c}{}		&	1.47	&	0.04	&	---	\\
112097	&	62933	&	4	&	6.237	&	0.000	&	+0.03	&	7330	&	95	&	+2.31	&	0.11	&		\multicolumn{2}{c}{}		&	0.97	&	0.04	&	$\delta$\,Sct	\\
113848	&	63948	&	4	&	6.014	&	0.000	&	+0.01	&	6620	&	49	&	+2.53	&	0.08	&		\multicolumn{2}{c}{}		&	0.89	&	0.03	&	---	\\
114879	&	64477	&	4	&	8.920	&	0.000	&	+0.02	&	7960	&	81	&		\multicolumn{2}{c}{}		&	+1.80	&	0.24	&	1.18	&	0.10	&	---	\\
114930	&	64525	&	4	&	9.017	&	0.000	&	+0.03	&	7080	&	73	&	+3.35	&	0.33	&	+2.36	&	0.14	&	0.96	&	0.06	&	---	\\
118623	&	66458	&	4	&	4.901	&	0.000	&	+0.03	&	7400	&	91	&	+0.98	&	0.07	&	+1.22	&	0.12	&	1.51	&	0.03	&	---	\\
119288	&	66860	&	4	&	6.162	&	0.000	&	+0.01	&	6560	&	57	&	+3.35	&	0.03	&		\multicolumn{2}{c}{}		&	0.56	&	0.01	&	other	\\
120500	&	67481	&	1	&	6.597	&	0.040	&	+0.02	&	8250	&	65	&	+0.62	&	0.18	&	+0.88	&	0.19	&	1.65	&	0.07	&	$\delta$\,Sct	\\
120896	&	67705	&	1	&	8.501	&	0.000	&	+0.03	&	7370	&	100&&&+1.56	&	0.16	&	1.27	&	0.06	&	$\delta$\,Sct	\\
123299	&	68756	&	4	&	3.644	&	0.003	&	$-$0.21	&	9790	&	362	&	$-$1.20	&	0.03	&		\multicolumn{2}{c}{}		&	2.38	&	0.01	&	other	\\
125162	&	69732	&	1	&	4.176	&	0.101	&	$-$0.04	&	8840	&	196	&	+1.66	&	0.01	&		\multicolumn{2}{c}{}		&	1.23	&	0.00	&	$\delta$\,Sct	\\
125489	&	70022	&	4	&	6.178	&	0.033	&	+0.02	&	7970	&	76	&	+2.14	&	0.05	&		\multicolumn{2}{c}{}		&	1.04	&	0.02	&	---	\\
125889	&	7294--813--1	&	2	&	9.835	&	0.126	&	+0.03	&	7320	&	100	&		\multicolumn{2}{c}{}		&	+2.24	&	0.28	&	1.00	&	0.11	&	const.	\\
128167	&	71284	&	4	&	4.467	&	0.000	&	+0.02	&	6720	&	45	&	+3.47	&	0.01	&		\multicolumn{2}{c}{}		&	0.51	&	0.00	&	---	\\
130158	&	72323	&	4	&	5.609	&	0.092	&	$-$0.46	&	10960	&	322	&	$-$1.27	&	0.17	&	$-$0.43	&	0.31	&	2.41	&	0.07	&	other	\\
130767	&	72505	&	1	&	6.903	&	0.000	&	$-$0.09	&	9170	&	234	&	+1.41	&	0.16	&	+1.36	&	0.09	&	1.36	&	0.04	&	const.	\\
138527	&	76069	&	2	&	6.217	&	0.140	&	$-$0.48	&	11080	&	200	&	+0.29	&	0.14	&		\multicolumn{2}{c}{}		&	1.79	&	0.05	&	---	\\
141569	&	77542	&	1	&	7.123	&	0.411	&	$-$0.27	&	10080	&	291	&	+1.39	&	0.15	&	+1.48	&	0.10	&	1.31	&	0.04	&	---	\\
141851	&	77660	&	4	&	5.090	&	0.000	&	+0.02	&	8100	&	125	&	+1.61	&	0.04	&		\multicolumn{2}{c}{}		&	1.26	&	0.01	&	const.	\\
142703	&	78078	&	1	&	6.113	&	0.000	&	+0.03	&	7200	&	89	&	+2.57	&	0.04	&		\multicolumn{2}{c}{}		&	0.87	&	0.02	&	$\delta$\,Sct	\\
142994	&	6791--1423--1	&	1	&	7.173	&	0.023	&	+0.03	&	6960	&	76	&		\multicolumn{2}{c}{}		&	+1.22	&	0.12	&	1.41	&	0.05	&	$\delta$\,Sct	\\
143148	&	7329--1376--1	&	4	&	7.388	&	0.012	&	+0.03	&	7180	&	85	&		\multicolumn{2}{c}{}		&	+1.96	&	0.08	&	1.12	&	0.03	&	const.	\\
144708	&	79005	&	4	&	5.756	&	0.201	&	$-$0.35	&	10460	&	397	&	+0.30	&	0.15	&		\multicolumn{2}{c}{}		&	1.78	&	0.06	&	---	\\
145782	&	79689	&	4	&	5.617	&	0.106	&	+0.02	&	8230	&	85	&	$-$0.05	&	0.08	&		\multicolumn{2}{c}{}		&	1.92	&	0.03	&	const.	\\
149130	&	81329	&	2	&	8.475	&	0.286	&	+0.03	&	6950	&	173	&	+1.44	&	0.53	&	+1.45	&	0.13	&	1.32	&	0.05	&	const.	\\
149303	&	80953	&	3	&	5.658	&	0.000	&	+0.02	&	8170	&	221	&	+1.45	&	0.09	&		\multicolumn{2}{c}{}		&	1.32	&	0.04	&	const.	\\
153747	&	83410	&	1	&	7.403	&	0.115	&	+0.02	&	8190	&	78	&	+1.01	&	0.25	&	+0.88	&	0.11	&	1.55	&	0.04	&	$\delta$\,Sct	\\
153808	&	83207	&	4	&	3.901	&	0.102	&	$-$0.34	&	10440	&	165	&	+0.41	&	0.01	&		\multicolumn{2}{c}{}		&	1.73	&	0.01	&	---	\\
154153	&	83650	&	2	&	6.184	&	0.335	&	+0.03	&	7400	&	567	&	+1.63	&	0.09	&		\multicolumn{2}{c}{}		&	1.25	&	0.04	&	const.	\\
156954	&	84895	&	3	&	7.670	&	0.000	&	+0.03	&	7070	&	61	&	+2.90	&	0.16	&	+2.90	&	0.07	&	0.74	&	0.03	&	const.	\\
159082	&	85826	&	4	&	6.416	&	0.184	&	$-$0.47	&	11020	&	199	&	+0.58	&	0.12	&	+0.62	&	0.18	&	1.67	&	0.05	&	---	\\
160928	&	86847	&	4	&	5.872	&	0.000	&	+0.02	&	7870	&	91	&	+1.58	&	0.06	&		\multicolumn{2}{c}{}		&	1.27	&	0.02	&	---	\\
161223	&	427--1650--1	&	1	&	7.443	&	0.404	&	+0.03	&	7430	&	107	&		\multicolumn{2}{c}{}		&	+0.32	&	0.16	&	1.77	&	0.06	&	$\delta$\,Sct	\\
\hline
\end{tabular}
\end{center}
\end{table*}

\addtocounter{table}{-1}

\begin{table*}
\begin{center}
\caption{continued.}
\begin{tabular}{cccrcrr@{ $\pm$ }rr@{ $\pm$ }rr@{ $\pm$ }rr@{ $\pm$ }rc}
\hline
\hline
HD & HIP/TYC & \hspace{-2mm}mem.\hspace{-2mm} & $m_\mathrm{V}$ & $A_\mathrm{V}$ & B.C. & \multicolumn{2}{c}{$T_\mathrm{eff}$} & \multicolumn{2}{c}{$M_\mathrm{V,Hip}$} & \multicolumn{2}{c}{$M_\mathrm{V,Gaia}$}  & \multicolumn{2}{c}{$\log L/L_{\sun}$}  & var? \\
& & & (mag) & (mag) & (mag) & \multicolumn{2}{c}{(K)} & \multicolumn{2}{c}{(mag)} & \multicolumn{2}{c}{(mag)} & \multicolumn{2}{c}{} & \\
\hline
161868	&	87108	&	4	&	3.740	&	0.052	&	$-$0.10	&	9210	&	151	&	+1.20	&	0.01	&		\multicolumn{2}{c}{}		&	1.42	&	0.01	&	---	\\
168740	&	90304	&	1	&	6.122	&	0.020	&	+0.03	&	7670	&	121	&	+1.84	&	0.05	&	+1.87	&	0.08	&	1.16	&	0.02	&	$\delta$\,Sct	\\
168947	&	7913--1173--1	&	1	&	8.114	&	0.124	&	+0.03	&	7510	&	101	&		\multicolumn{2}{c}{}		&	+1.10	&	0.18	&	1.46	&	0.07	&	$\delta$\,Sct	\\
169009	&	90083	&	4	&	6.322	&	0.665	&	$-$0.27	&	10120	&	441	&	+0.60	&	0.30	&	+1.46	&	0.11	&	1.32	&	0.04	&	---	\\
169022	&	90185	&	4	&	1.802	&	0.112	&	$-$0.18	&	9640	&	125	&	$-$1.52	&	0.02	&		\multicolumn{2}{c}{}		&	2.51	&	0.01	&	---	\\
169142	&	6856--876--1	&	3	&	8.152	&	0.000	&	+0.02	&	6700	&	322	&		\multicolumn{2}{c}{}		&	+2.81	&	0.07	&	0.78	&	0.03	&	---	\\
170000	&	89908	&	4	&	4.223	&	0.009	&	$-$0.77	&	12430	&	139	&	$-$0.62	&	0.08	&		\multicolumn{2}{c}{}		&	2.15	&	0.03	&	other	\\
170680	&	90806	&	1	&	5.120	&	0.107	&	$-$0.22	&	9860	&	302	&	+0.84	&	0.04	&		\multicolumn{2}{c}{}		&	1.56	&	0.02	&	---	\\
172167	&	91262	&	2	&	0.074	&	0.042	&	$-$0.13	&	9400	&	384	&	+0.61	&	0.01	&		\multicolumn{2}{c}{}		&	1.66	&	0.00	&	$\delta$\,Sct	\\
174005	&	92296	&	1	&	6.493	&	0.230	&	+0.03	&	7790	&	138	&	$-$0.56	&	0.27	&	+0.15	&	0.12	&	1.84	&	0.05	&	---	\\
175445	&	92884	&	4	&	7.781	&	0.129	&	$-$0.07	&	9060	&	333	&	+1.50	&	0.25	&		\multicolumn{2}{c}{}		&	1.30	&	0.10	&	---	\\
177756	&	93805	&	4	&	3.427	&	0.000	&	$-$0.57	&	11460	&	158	&	+0.53	&	0.05	&		\multicolumn{2}{c}{}		&	1.69	&	0.02	&	---	\\
179791	&	94478	&	4	&	6.472	&	0.000	&	+0.02	&	8090	&	118	&	+0.62	&	0.27	&	+0.20	&	0.22	&	1.82	&	0.09	&	const.	\\
181470	&	94932	&	4	&	6.243	&	0.062	&	$-$0.26	&	10050	&	186	&	$-$0.22	&	0.13	&	+0.37	&	0.27	&	1.99	&	0.05	&	---	\\
183007	&	95823	&	3	&	5.702	&	0.000	&	+0.03	&	7670	&	106	&	+1.52	&	0.05	&		\multicolumn{2}{c}{}		&	1.21	&	0.02	&	---	\\
183324	&	95793	&	1	&	5.783	&	0.330	&	$-$0.22	&	9840	&	150	&	+1.52	&	0.05	&		\multicolumn{2}{c}{}		&	1.29	&	0.02	&	$\delta$\,Sct	\\
184190	&	7936--1366--1	&	4	&	9.737	&	0.138	&	+0.03	&	7160	&	95	&		\multicolumn{2}{c}{}		&	+2.71	&	0.15	&	0.82	&	0.06	&	---	\\
184779	&	7944--1500--1	&	1	&	8.911	&	0.000	&	+0.03	&	7130	&	81	&		\multicolumn{2}{c}{}		&	+2.06	&	0.14	&	1.08	&	0.06	&	const.	\\
187949	&	97849	&	4	&	6.481	&	0.000	&	+0.02	&	7970	&	99	&	+1.10	&	0.15	&		\multicolumn{2}{c}{}		&	1.46	&	0.06	&	other	\\
188164	&	98346	&	1	&	6.373	&	0.077	&	+0.02	&	8160	&	64	&	+1.18	&	0.11	&	+1.28	&	0.15	&	1.39	&	0.06	&	const.	\\
188728	&	98103	&	4	&	5.285	&	0.047	&	$-$0.21	&	9810	&	155	&	+1.10	&	0.04	&		\multicolumn{2}{c}{}		&	1.46	&	0.01	&	---	\\
191850	&	8392--2637--1	&	1	&	9.682	&	0.022	&	+0.03	&	7370	&	95	&		\multicolumn{2}{c}{}		&	+1.75	&	0.29	&	1.20	&	0.12	&	$\delta$\,Sct	\\
192640	&	99770	&	1	&	4.946	&	0.043	&	+0.02	&	7960	&	92	&	+1.75	&	0.02	&		\multicolumn{2}{c}{}		&	1.20	&	0.01	&	$\delta$\,Sct	\\
193281	&	100288	&	3	&	6.629	&	0.100	&	+0.02	&	8020	&	72	&		\multicolumn{2}{c}{}		&	+0.51	&	0.18	&	1.70	&	0.07	&	const.	\\
196821	&	101919	&	4	&	6.074	&	0.000	&	$-$0.30	&	10250	&	58	&	$-$0.64	&	0.19	&		\multicolumn{2}{c}{}		&	2.16	&	0.07	&	---	\\
198160	&	102962	&	1	&	6.173	&	0.018	&	+0.02	&	7930	&	77	&	+1.74	&	0.11	&		\multicolumn{2}{c}{}		&	1.20	&	0.04	&	---	\\
198161	&	102962	&	1	&	6.518	&	0.018	&	+0.02	&	7900	&	77	&	+2.09	&	0.11	&		\multicolumn{2}{c}{}		&	1.07	&	0.04	&	---	\\
200841	&	104042	&	4	&	8.270	&	0.362	&	$-$0.26	&	10070	&	319	&		\multicolumn{2}{c}{}		&	+0.18	&	0.24	&	1.83	&	0.10	&	---	\\
201019	&	104408	&	3	&	8.380	&	0.167	&	+0.02	&	6660	&	257	&	+2.27	&	0.31	&	+1.63	&	0.15	&	1.25	&	0.06	&	---	\\
201184	&	104365	&	4	&	5.307	&	0.072	&	$-$0.25	&	9990	&	189	&	+1.53	&	0.04	&		\multicolumn{2}{c}{}		&	1.29	&	0.01	&	---	\\
204041	&	105819	&	1	&	6.454	&	0.034	&	+0.02	&	8090	&	74	&	+2.07	&	0.10	&	+2.10	&	0.05	&	1.06	&	0.02	&	const.	\\
204754	&	106052	&	4	&	6.135	&	0.666	&	$-$0.80	&	12550	&	108	&	$-$2.49	&	0.25	&	$-$1.86	&	0.31	&	2.90	&	0.10	&	---	\\
204965	&	106171	&	4	&	6.011	&	0.060	&	+0.01	&	8420	&	131	&	$-$0.05	&	0.09	&	+0.15	&	0.13	&	1.92	&	0.04	&	---	\\
207978	&	107975	&	4	&	5.524	&	0.053	&	$-$0.01	&	6330	&	99	&	+3.28	&	0.02	&		\multicolumn{2}{c}{}		&	0.59	&	0.01	&	---	\\
210111	&	109306	&	1	&	6.375	&	0.000	&	+0.03	&	7470	&	103	&	+1.91	&	0.09	&	+1.92	&	0.04	&	1.13	&	0.02	&	$\delta$\,Sct	\\
210418	&	109427	&	4	&	3.520	&	0.084	&	$-$0.03	&	8730	&	226	&	+1.18	&	0.05	&		\multicolumn{2}{c}{}		&	1.43	&	0.02	&	---	\\
212061	&	110395	&	4	&	3.844	&	0.006	&	$-$0.30	&	10250	&	256	&	+0.33	&	0.11	&		\multicolumn{2}{c}{}		&	1.77	&	0.05	&	---	\\
212150	&	110116	&	4	&	6.612	&	0.215	&	$-$0.24	&	9940	&	377	&	$-$0.60	&	0.27	&	$-$0.21	&	0.19	&	1.98	&	0.08	&	---	\\
213669	&	111411	&	2	&	7.409	&	0.000	&	+0.03	&	7350	&	124	&	+2.24	&	0.16	&	+2.10	&	0.09	&	1.06	&	0.04	&	$\delta$\,Sct	\\
214454	&	111674	&	4	&	4.644	&	0.000	&	+0.03	&	7380	&	113	&	+1.04	&	0.02	&		\multicolumn{2}{c}{}		&	1.48	&	0.01	&	---	\\
216847	&	113351	&	2	&	7.059	&	0.006	&	+0.03	&	7340	&	97	&	+1.16	&	0.16	&	+1.20	&	0.13	&	1.42	&	0.05	&	const.	\\
217782	&	113788	&	4	&	5.085	&	0.281	&	$-$0.08	&	9140	&	144	&	$-$0.75	&	0.14	&		\multicolumn{2}{c}{}		&	2.20	&	0.06	&	other	\\
218396	&	114189	&	1	&	5.953	&	0.000	&	+0.03	&	7230	&	95	&	+2.98	&	0.06	&	+2.92	&	0.06	&	0.73	&	0.02	&	$\gamma$\,Dor	\\
220061	&	115250	&	3	&	4.581	&	0.000	&	+0.03	&	7700	&	87	&	+1.10	&	0.04	&		\multicolumn{2}{c}{}		&	1.46	&	0.02	&	$\delta$\,Sct	\\
220278	&	115404	&	3	&	5.441	&	0.000	&	+0.03	&	7610	&	101	&	+1.36	&	0.11	&		\multicolumn{2}{c}{}		&	1.35	&	0.04	&	---	\\
221756	&	116354	&	1	&	5.557	&	0.093	&	$-$0.01	&	8630	&	200	&	+0.94	&	0.05	&		\multicolumn{2}{c}{}		&	1.52	&	0.02	&	$\delta$\,Sct	\\
222303	&	4292--245--1	&	3	&	9.159	&	1.001	&	+0.03	&	6960	&	69	&		\multicolumn{2}{c}{}		&	+0.17	&	0.21	&	1.83	&	0.09	&	---	\\
223352	&	117452	&	2	&	4.572	&	0.062	&	$-$0.23	&	9930	&	197	&	+1.39	&	0.02	&		\multicolumn{2}{c}{}		&	1.35	&	0.01	&	---	\\
228509	&	3151--1876--1	&	3	&	9.237	&	0.105	&	+0.03	&	7220	&	50	&		\multicolumn{2}{c}{}		&	+0.68	&	0.36	&	1.63	&	0.14	&	---	\\
290799	&	4767--765--1	&	1	&	10.703	&	0.022	&	+0.02	&	7940	&	71	&		\multicolumn{2}{c}{}		&	+3.10	&	0.25	&	0.66	&	0.10	&	$\delta$\,Sct	\\
294253	&	4770--1225--1	&	1	&	9.645	&	0.228	&	$-$0.37	&	10580	&	211	&		\multicolumn{2}{c}{}		&	+1.49	&	0.28	&	1.30	&	0.11	&	const.	\\
\hline
\end{tabular}
\end{center}
\end{table*}

\section{The HR Diagram}
\label{sec:HRD}

\begin{figure*}
\begin{center}
\includegraphics[width=0.98\textwidth]{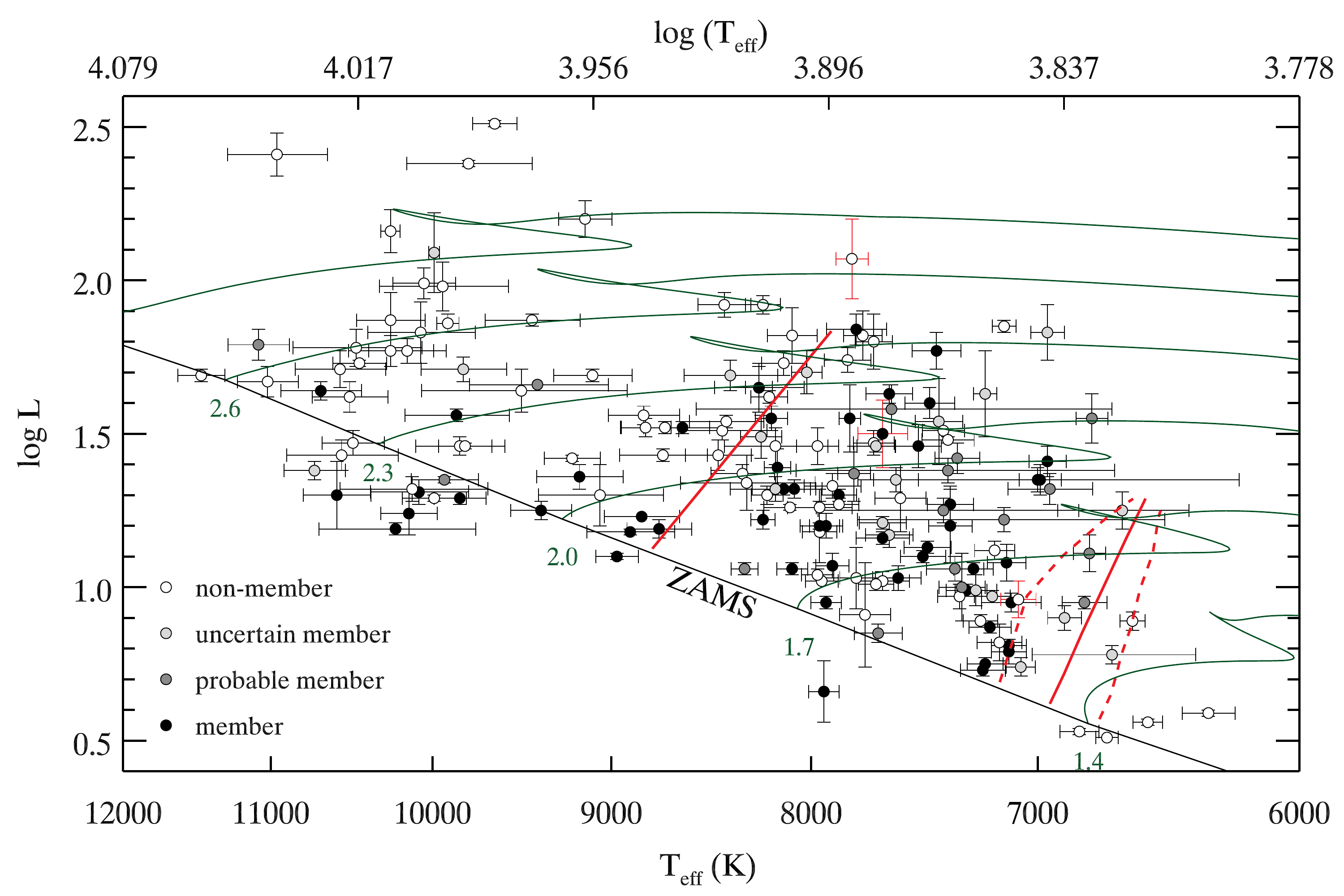}
\caption{An HR-diagram of 172 $\lambda$\,Boo stars and candidates with Hipparcos or Gaia parallaxes. 
The probability of membership in the $\lambda$\,Boo class is indicated by symbol shade. 
Evolutionary tracks at $Z=0.02$ are shown as green lines, 
with their corresponding masses (in M$_{\odot}$) written beneath the ZAMS (black line). 
The $\delta$\,Sct and $\gamma$\,Dor instability strips are delimited with solid and dashed red lines, respectively. The temperature scale is logarithmic. The three stars with anomalously large $\Delta \overline{M_\mathrm{V}}$ are shown with red error bars.}
\label{fig:HRD}
\end{center}
\end{figure*}

Fig.\,\ref{fig:HRD} shows the selected $\lambda$\,Boo stars and candidates on an HR-diagram, colour-coded by membership status. 
We have added evolutionary tracks at $Z=0.02$,
calculated with time-dependent convection (TDC) models of mixing length 
$\alpha_{\rm MLT} = 1.8$\,$h_{\rm p}$ (pressure scale heights) from \citet{grigahceneetal2005}. This value of $\alpha_{\rm MLT}$ was empirically validated by matching the observed and theoretical instability strips for $\delta$\,Sct and $\gamma$\,Dor stars \citep{dupretetal2005b}. We have included those same theoretical instability strips, which were computed for the radial modes for the $\delta$\,Sct stars, and $\ell=1$ modes for the $\gamma$\,Dor stars.

The global metallicity of these models, $Z$, is higher than the measured metallicity of $\lambda$\,Boo stars, 
which deserves some discussion. The chemical peculiarities of $\lambda$\,Boo stars are believed to be limited 
to the surface, so TDC models with a global metallicity typical of A stars in the galactic plane were chosen. 
Generally, metal-poor stars are hotter and less luminous than metal-rich stars of the same mass and age due to 
differences in the opacity. While we expect no large systematic displacement of the $\lambda$\,Boo stars in 
Fig.\,\ref{fig:HRD} with respect to the evolutionary tracks, any displacement of these stars that does occur 
will be to the lower left. This is one potential explanation of some objects that lie below the ZAMS. Other 
explanations include underestimated extinction, particularly because objects at the ZAMS are located in 
or near star formation regions (e.g.\ two objects that are in the Orion OB1 association), or technical problems 
with the Gaia parallaxes in the first data release due to undetected binaries \citep{lindgreenetal2016}.

\subsection{Influence on pulsation}
\label{ssec:pulsation}

The fact that the low surface metallicity is attributed to the accretion of material with a low dust-to-gas mass ratio 
has strong implications on the driving of pulsation. The pressure-mode ($p$~mode) oscillations of $\delta$\,Sct stars 
are driven by the opacity mechanism operating on the He\,{\sc ii} partial ionisation zone. \citet{murphy2014} argued 
that the delivery of fresh helium at the surface, combined with the typically rapid rotation of $\lambda$\,Boo stars, 
acts to prevent helium from gravitationally settling in these stars. The driving zone is therefore well-stocked with 
helium, which \citet{murphy2014} argued would lead to stronger oscillation amplitudes in these stars, and/or a higher 
fraction of them being pulsators, compared to an otherwise identical sample of chemically normal stars. We have therefore collated pulsation properties from the literature 
\citep{antonello&mantegazza1982, paunzenetal1997b, paunzenetal1998b, paunzenetal2002a, paunzenetal2014a, paunzenetal2015}.

\begin{figure*}
\begin{center}
\includegraphics[width=0.98\textwidth]{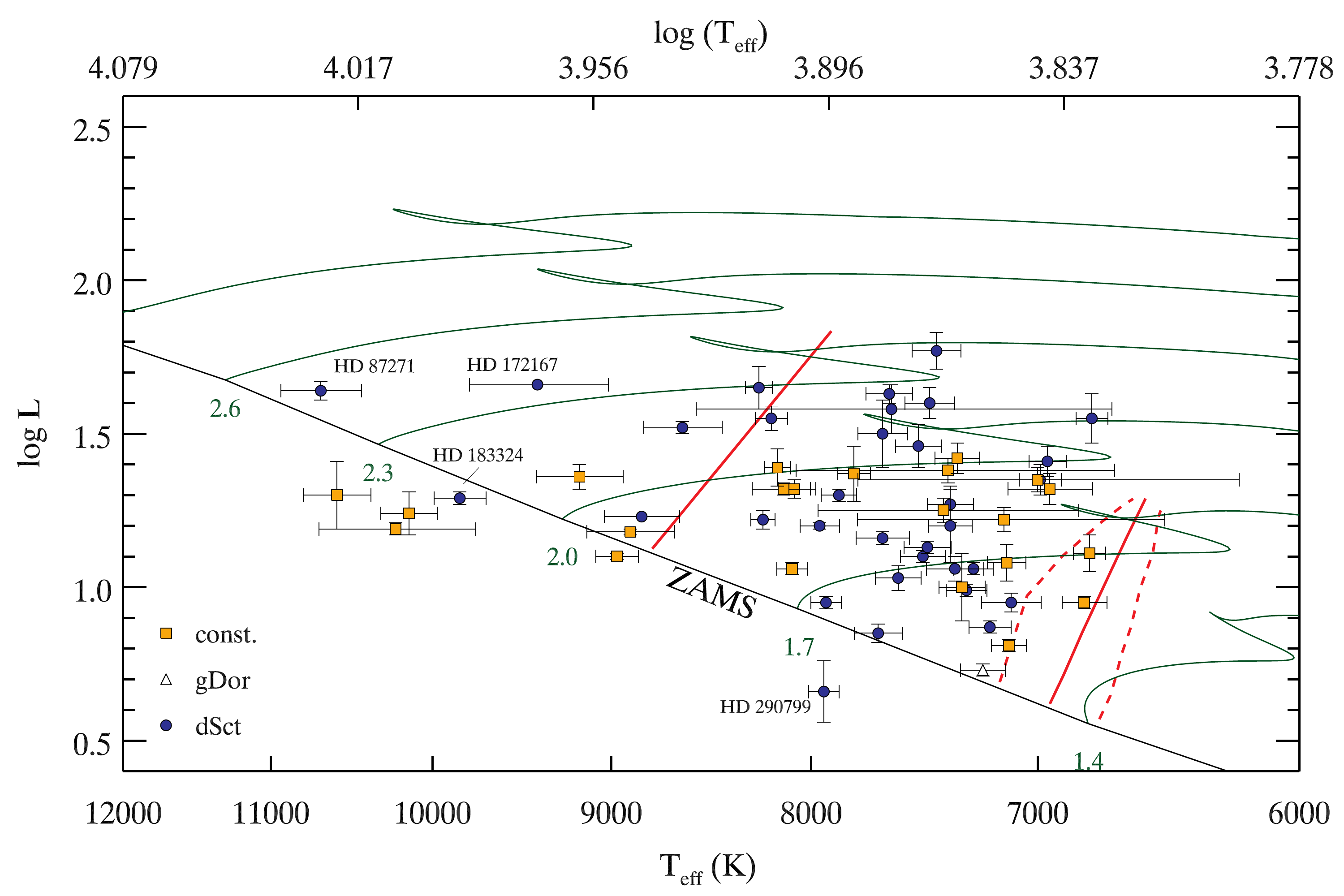}
\caption{As Fig.\,\ref{fig:HRD}, but showing only the 52 members and 17 probable members of the $\lambda$\,Boo class. Different 
symbol types show the pulsation properties. Constant stars show no variability greater than a few mmag. The one $\gamma$\,Dor 
star is HR\,8799 \citep{gray&kaye1999}. Stars labelled with HD numbers are discussed in the text.}
\label{fig:puls}
\end{center}
\end{figure*}

Fig.\,\ref{fig:puls} shows the pulsation properties of members and probable members of the $\lambda$\,Boo class. There 
are 16 (38\%) non-pulsating $\lambda$\,Boo stars in the instability strip, at the photometric precision achieved 
in ground-based surveys (typically a few mmag). \textit{Kepler} observations have shown that many $\delta$\,Sct stars 
have oscillation amplitudes below 1\,mmag, so these objects are readily explained as variables for which 
the photometry was insufficiently sensitive to detect the oscillations. Three pulsating stars are much hotter than outside the instability
strip boundary. In the following, the characteristics of these objects plus the $\delta$\,Sct star well below the ZAMS are discussed in more detail.

{\it HD\,87271:} The Str{\"o}mgren-Crawford indices are the following: $(b-y)$\,=\,+0.151\,mag, $m_\mathrm{1}$\,=\,+0.094\,mag,	
$c_\mathrm{1}$\,=\,+0.939\,mag, and $\beta$\,=\,2.775\,mag. Those are not self-consistent \citep{crawford1978,crawford1979}.
If we assume that the reddening is zero, leading to $T_{\rm eff} \sim 7500$\,K, then the $c_\mathrm{1}$ index is much too high for a MS star. 
On the other hand, the $\beta$ index is not consistent with a hotter MS object. A 
detailed spectroscopic analysis is required for definitive characterisation. 

{\it HD\,172167 (Vega):} The variability of this bright standard star has been extensively reviewed by \citet{butkovskaya2014}.
It seems to vary on time scales typical for classical pulsation up to several decades indicating similarities to the Solar cycle.
The membership to the $\lambda$\,Boo group has been extensively discussed by \citet{murphyetal2015b}.

{\it HD\,183324:} Spectroscopically determined 
values of $T_\mathrm{eff}$ \citep{heiteretal1998} are in good agreement with our photometrically derived
one. It has a very short period (30 minutes) and a very low surface metallicity
($Z=-1.5$\,dex). Therefore, HD\,183324 seems to be quite outstanding compared to the other pulsating $\lambda$\,Boo stars.
\citet{gerbaldietal2003} reported that the radial velocity is variable but they were not able to detect any optical companion 
using speckle interferometry. This variability could be caused, other than by pulsation, by undetected binarity, which would explain
the location below the ZAMS.

{\it HD\,290799:} This star is a member of the Orion OB1 association \citep{gray&corbally1993}. The estimated reddening
of 0.022\,mag is low for this region. \citet{paunzenetal2002a} estimated $M_\mathrm{V}$\,=\,+2.62$\pm$0.30\,mag on
the basis of photometric data and the distance to the Orion OB1 association, only. This value is about 0.5\,mag brighter
than our estimate and would put HD\,290799 on the ZAMS. There are no Geneva colours available, which prevents us 
from checking the intrinsic consistency of the Str{\"o}mgren-Crawford ones. Therefore, we are not able to resolve the inconsistency for
HD\,290799.

Asteroseismology is a promising tool to assess whether the metal depletion in $\lambda$\,Boo stars is global or limited to the surface instead. 
A detailed analysis with ultra-precise \textit{Kepler} (and later, TESS) light curves is highly warranted, and will be the subject of future work.

\section{Discussion}
\label{sec:discussion}

The distribution of $\lambda$\,Boo stars on the HR-diagram appears divided loosely around the classical ($\delta$\,Sct) instability 
strip. On the upper main sequence, which we define as the region hotter than the blue edge of the instability strip, the 
$\lambda$\,Boo stars are confined to the ZAMS. However, within the instability strip, the $\lambda$\,Boo stars occupy a 
wide range of MS ages, with some possibly above the TAMS. Redward of the instability strip, no $\lambda$\,Boo stars are observed.

It it not clear whether this distribution is to be explained in terms of the physics operating at different effective temperatures, or at different masses instead. 
Its division around the instability strip is coincidental, particularly since not all group members are observed to pulsate, and the $\delta$\,Sct hot edge is clearly not a hard boundary beyond which all $\lambda$\,Boo stars are confined to the ZAMS.

In the coolest stars, the weaker radiation field is least efficient at preventing the accretion of dust, 
so the composition of any accreted material is not highly anomalous to begin with. 
Further, the surface convection zone is deep and massive enough that any accreted material is thoroughly mixed 
and a large contamination is required before peculiarities are observed.

In the hottest stars the radiation field is strongest, and the quantity of material required to pollute the surface is lower, but neither of these principles has resulted in more $\lambda$\,Boo stars towards the TAMS.
Thus the confinement of $\lambda$\,Boo stars to the ZAMS indicates that the 
phenomenon is associated with star formation in some way. In what follows, we reconcile the observations with current theory.

There are several different hypotheses surrounding the development of $\lambda$\,Boo peculiarities, each with their own merits. All of these hypotheses are related to dust-gas separation and the accretion of dust-depleted material, with two notable exceptions: ({\sc i}) the refuted composite-spectrum hypothesis, which was covered in Sect.\,\ref{sec:intro}, and has been more comprehensively reviewed by \citet{murphyetal2015b}; and ({\sc ii}) the diffusion and mass-loss mechanism \citep{michaud&charland1986}, which tends to produce overabundances of Fe-peak elements typical of Am stars \citep{vicketal2010,vicketal2011}, rather than the underabundances observed in $\lambda$\,Boo stars. We suggest that it is not a single mechanism that is responsible, but that several channels can form $\lambda$\,Boo spectra under certain conditions.

We start with the hot ZAMS $\lambda$\,Boo stars. Approximately half of the Herbig Ae/Be (HAeBe) stars show abundance patterns in common with $\lambda$\,Boo stars \citep{folsometal2012}. It is possible that stars begin their main-sequence lives with this chemical imprint of star formation, resulting from partial accretion of the protoplanetary disk. \citet{kamaetal2015} suggested that jovian planets can exaggerate the process by trapping dust in the disk. Particularly for massive stars the protoplanetary disk is short-lived \citep{ribasetal2015}, and rapid rotation mixes these stars on short timescales \citep[e.g.][]{maeder&meynet2000,espinosa&rieutord2013}, so the peculiarities disappear shortly after the stars begin hydrogen fusion, therefore the phenomenon is limited to the ZAMS.

At lower masses, the $\lambda$\,Boo phenomenon is observed across the whole main-sequence phase, so the accretion of protoplanetary material cannot be the sole explanation. An alternative is that the gas-rich material causing the peculiarities has been ablated from hot Jupiters \citep{jura2015}. The idea has support from observations that A stars are efficient at forming high-mass planets \citep{johnsonetal2011,murphyetal2016c}, and some are observed in small orbits \citep{colliercameronetal2010}. The trapping of material in a planetary reservoir allows the accretion episode to be delayed, which accounts for the older $\lambda$\,Boo stars. The rather abrupt termination in $\lambda$\,Boo stars as a function of mass, seen in Fig.\,\ref{fig:HRD}, is also explained by this mechanism. The peak of the planet occurrence 
rate distribution, as a function of host-star mass, has a sharp cut-off at the high mass end: $1.9^{+0.1}_{-0.5}$\,M$_{\odot}$ \citep{reffertetal2015}. Only around 23\% of $\lambda$\,Boo stars have an infrared excess \citep{paunzenetal2003}, so a further strength of this mechanism is that it does not demand one.

The hot-Jupiter hypothesis is not perfect -- it cannot explain the lack of $\lambda$\,Boo stars in certain cluster environments. \citet{gray&corbally2002} assessed 220 stars in intermediate-age (35--830\,Myr) clusters, for which cluster membership was certain and no bias for or against the detection of $\lambda$\,Boo peculiarity was known. Not a single $\lambda$\,Boo star was found. For {\it field} $\lambda$\,Boo stars, they estimated the occurrence rate to be 2.0--2.5\% of the population. Applying a binomial distribution to their results for clusters at the field occurrence rate of 2.0\%, we find the probability of detecting zero $\lambda$\,Boo stars, P(X=0) = 0.012. This statistically significant result is not explained by the hot-Jupiter hypothesis, given that such planets have no difficulty forming in clusters \citep{lovis&mayor2007,satoetal2007,quinnetal2012,meibometal2013}.

The lack of $\lambda$\,Boo stars in intermediate-age clusters is explained if the dominant source of accreted material is diffuse ISM clouds \citep{martinez-galarzaetal2009}, because such clouds are not observed in clusters. The diffuse ISM hypothesis is also capable of explaining the lack of $\lambda$\,Boo stars in the later main-sequence phases of more massive stars, since their shorter lifetimes reduce the chances of encountering an ISM cloud after leaving their stellar nursery. In fact, none of the aforementioned observations are incompatible with the ISM accretion model, but it is not without problems. The star $\delta$\,Vel has been presented as a cast-iron example of an object interacting with the ISM but it does not belong to the $\lambda$\,Boo class \citep{gasparetal2008}. Confirmed interactions with the ISM are rare, particularly in the local bubble where most $\lambda$\,Boo stars are found but interstellar gas is deficient \citep{lallementetal2003}, whereas several A stars with resolved debris disks are known \citep{boothetal2013}. While \citet{martinez-galarzaetal2009} conceded that their models of infrared excesses for A stars fitted debris disks just as well as the interactions with the ISM they were promoting, \citet{draperetal2016} have found infrared emission around 9 $\lambda$\,Boo stars that is {\it only} consistent with debris disks. There are also observational concerns that the ISM accretion rates required for $\lambda$\,Boo peculiarities are unachievable \citep{jura2015}.

Finally, we note that there are many more $\lambda$\,Boo stars on the ZAMS for the hotter half of the sample in Fig.\,\ref{fig:HRD} than for the cooler half. This is explained by selection effects. For instance, pre-main-sequence stars have often been targets in the search for $\lambda$\,Boo stars \citep{acke&waelkens2004,folsometal2012}, and the majority of known pre-main-sequence A stars lie at the hotter end of the temperature range of Fig.\,\ref{fig:HRD}. The latter is easily seen in catalogues of stellar spectral types. Of the 150 stars with spectral types of \mbox{A[0-9]e} in the \citet{skiff2014} catalogue, 102 (68\%) have spectral types of A2e or earlier.

\section{Conclusions}
\label{sec:conclusions}

The location of $\lambda$\,Boo stars on an HR diagram suggests there is more than one channel by which accretion may cause $\lambda$\,Boo peculiarities in stellar spectra. We propose this as the reason that a single formation channel has not been agreed upon after decades of attempts. The relative importance of the different channels varies with age, temperature and environment (cluster vs. field).

Young stars form $\lambda$\,Boo-like spectra via accretion of a protoplanetary disk, with dust-gas separation being more efficient if giant planets are present. Older stars may accrete gas ablated from hot Jupiters, but the absence of $\lambda$\,Boo stars in intermediate-age clusters suggests this cannot be the sole mechanism forming these peculiarities. Interaction with diffuse clouds in the ISM has strong theoretical basis, but is also not the dominant mechanism according to observations of debris disks around $\lambda$\,Boo stars, and of ISM accretion rates onto other stars.

The future of $\lambda$\,Boo research is divided between further observations and computative analyses. Quantitative modelling, considering all channels forming $\lambda$\,Boo-like spectra, is highly sought after. A homogeneous investigation of the infrared properties of class members can further pin down the source of dust. And finally, asteroseismology will probe the penetrative depth of the metal depletion.


\section*{Acknowledgements}
We thank J.~Molenda-{\.Z}akowicz for discussion on the paper. This research was supported by the Australian Research Council. Funding for the Stellar Astrophysics Centre is provided by The 
Danish National Research Foundation (grant agreement no.: DNRF106). This work has made use of data from the European Space Agency 
(ESA) mission {\it Gaia} (\url{http://www.cosmos.esa.int/gaia}), processed by the {\it Gaia} Data Processing and Analysis 
Consortium (DPAC, \url{http://www.cosmos.esa.int/web/gaia/dpac/consortium}). Funding for the DPAC has been provided by national 
institutions, in particular the institutions participating in the {\it Gaia} Multilateral Agreement.
This research has made use of the Washington Double Star Catalog maintained at the U.S. Naval Observatory.

\bibliographystyle{mnras}
\bibliography{sjm_bibliography}


\end{document}